\documentclass[conference]{IEEEtran}

\IEEEoverridecommandlockouts
\usepackage{cite}
\usepackage{amsmath,amssymb,amsfonts}
\usepackage{graphicx}
\usepackage{textcomp}
\usepackage{utfsym}
\usepackage{listings}
\usepackage{algorithm}
\usepackage{algorithmicx}
\usepackage[noend]{algpseudocode}
\usepackage{array}
\usepackage{multirow}
\usepackage{pifont}
\usepackage{afterpage}
\usepackage[normalem]{ulem}
\usepackage{xspace}
\usepackage{enumitem}
\usepackage{subfig}
\usepackage[most]{tcolorbox}
\usepackage{xcolor}
\definecolor{codegray}{rgb}{0.5,0.5,0.5}
\definecolor{codegreen}{rgb}{0,0.6,0}
\definecolor{codepurple}{rgb}{0.58,0,0.82}
\definecolor{backcolour}{rgb}{0.95,0.95,0.92}
\usepackage{booktabs}
\usepackage{tabularx}
\usepackage{float}
\usepackage{url}
\usepackage{hyperref}  
\usepackage{cleveref}  

\def\BibTeX{{\rm B\kern-.05em{\sc i\kern-.025em b}\kern-.08em
    T\kern-.1667em\lower.7ex\hbox{E}\kern-.125emX}}

\tcbset{
  colback=gray!15,    
  colframe=gray!20,   
  arc=1mm,            
  boxrule=0.5pt,      
  width=\linewidth,   
  left=5pt,           
  right=5pt,          
  top=5pt,            
  bottom=5pt          
}
\newcommand{\xiaoxue}[1]{\textcolor{orange}{\textbf{Xiaoxue}: #1}}
\newcommand{\likai}[1]{\textcolor{blue}{\textbf{Likai}: #1}}

\newcommand{\tool}{\textsc{SMCFixer}\xspace}

\newcommand{\retriever}{\textsc{SocR}\xspace} 

\newcommand{\dataseta}{\textsc{Dataset-A}\xspace} 
\newcommand{\datasetb}{\textsc{Dataset-B}\xspace} 
\newcommand{\datasetc}{\textsc{Dataset-C}\xspace} 

\begin{document}

\title{Bridging Solidity Evolution Gaps: An LLM-Enhanced Approach for Smart Contract Compilation Error Resolution
\thanks{\textsuperscript{*}Corresponding Author: Xiaoxue Ren }
}

\author{
\IEEEauthorblockN{Likai Ye\IEEEauthorrefmark{1},
Mengliang Li\IEEEauthorrefmark{1},
Dehai Zhao\IEEEauthorrefmark{2},
Jiamou Sun\IEEEauthorrefmark{3},
Xiaoxue Ren\IEEEauthorrefmark{1}}

\IEEEauthorblockA{\IEEEauthorrefmark{1}Hangzhou High-Tech Zone (Binjiang) Institute of Blockchain and Data Security, Zhejiang University, China\\
\IEEEauthorrefmark{2}School of Software Technology, Zhejiang University, China\\
\IEEEauthorrefmark{3}CSIRO's Data61, Australia\\
Email:yelikai@zju.edu.cn,123lmliang@zju.edu.cn,zhaodh.upc@gmail.com,u587153@anu.edu.au,xxren@zju.edu.cn}


}


\maketitle

\begin{abstract}
Solidity, the dominant smart contract language for Ethereum, has rapidly evolved with frequent version updates to enhance security, functionality, and developer experience. However, these continual changes introduce significant challenges, particularly in compilation errors, code migration, and maintenance. Therefore, we conduct an empirical study to investigate the challenges in the Solidity version evolution and reveal that 81.68\% of examined contracts encounter errors when compiled across different versions, with 86.92\% of compilation errors.

To mitigate these challenges, we conducted a systematic evaluation of large language models (LLMs) for resolving Solidity compilation errors during version migrations. Our empirical analysis across both open-source (LLaMA3, DeepSeek) and closed-source (GPT-4o, GPT-3.5-turbo) LLMs reveals that although these models exhibit error repair capabilities, their effectiveness diminishes significantly for semantic-level issues and shows strong dependency on prompt engineering strategies. This underscores the critical need for domain-specific adaptation in developing reliable LLM-based repair systems for smart contracts.

Building upon these insights, we introduce \tool, a novel framework that systematically integrates expert knowledge retrieval with LLM-based repair mechanisms for Solidity compilation error resolution. The architecture comprises three core phases: (1) context-aware code slicing that extracts relevant error information; (2) expert knowledge retrieval from official documentation; and (3) iterative patch generation for Solidity migration.
Experimental validation across Solidity version migrations demonstrates our approach's statistically significant 24.24\% improvement over baseline GPT-4o on real-world datasets, achieving near-perfect 96.97\% accuracy.


\end{abstract}


\section{Introduction}
Solidity is a high-level programming language specifically designed for writing smart contracts on blockchain platforms, primarily Ethereum~\cite{tikhomirov2018ethereum,li2023convmhsa}.
Smart contracts are self-executing code segments that enforce and execute agreements when predefined conditions are met~\cite{wood2014ethereum}. 
As a key enabler of decentralized applications (DApps) and blockchain-based transaction automation, Solidity plays a crucial role in the functionality of blockchain ecosystems~\cite{sillaber2017life}.
Over time, Solidity has evolved 84 times significantly from \emph{v0.4.1} to \emph{v0.8.23}, with major updates improving security, functionality, and developer experience~\cite{Mitropoulos2024}. 
For instance, early versions focused on core language features, while later versions introduced breaking changes to enhance clarity and robustness, such as explicitness requirements, overflow checks, and new syntax for modularity. 
This evolution reflects its ongoing commitment to addressing security concerns and optimizing performance in response to the growing complexity of blockchain use cases~\cite{bartoletti2017empirical}.

Although Solidity's frequent evolution has improved the language's security and functionality, it has also brought many coding challenges to developers, as they need to constantly learn and adapt to these changes to ensure code usability. 
Especially when a project relies on multiple smart contracts or requires long-term maintenance, compatibility issues between different Solidity versions may cause the code to be unavailable or need to be rewritten. 

\begin{figure*}[htbp] 
\centering
\includegraphics[width=\textwidth]{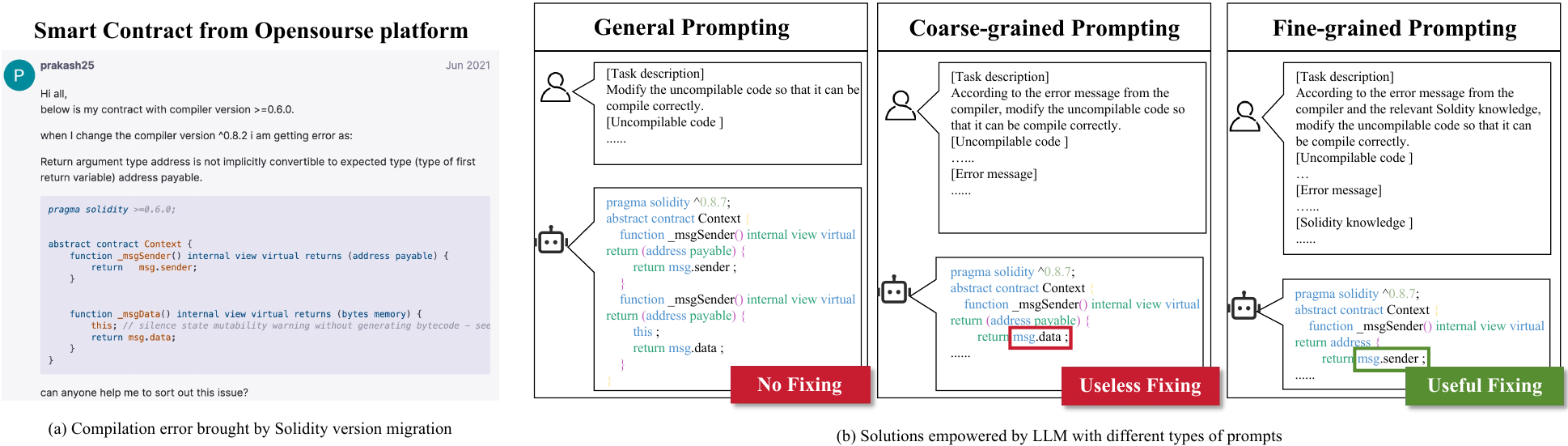} 
\caption{An Example for Investigating LLM's Capability of Fixing Solidity Compilation Error}
\label{fig:motivating_example} 
\end{figure*}

For example, Fig.~\ref{fig:motivating_example}-(a) illustrates a type error of the Solidity smart contract caused by a version migration from \emph{v0.6.x} to \emph{v0.8.x}, as discussed in a Stack Overflow post.
In this case, the type of \texttt{msg.sender} is \texttt{address}, while the function declaration specifies a return type of \texttt{address payable}.
According to official materials~\cite{solidity080}, in Solidity \emph{v0.6.x}, implicit conversion between \texttt{address} and \texttt{address payable} is allowed.
However, in \emph{v0.8.x}, the conversion rules for these types have changed—implicit conversion is no longer permitted, and explicit conversion is now required.
Consequently, without awareness of these language evolution changes, migrating a contract from \emph{v0.6.x} to \emph{v0.8.x} can lead to a type error during compilation.

However, as far as we know, there is no related work investigating the challenges and corresponding solutions brought by the frequent evolution of the Solidity language.
To fill this knowledge gap, we first conduct an empirical study on the official materials to investigate ``\emph{RQ1: What is the current state of the Solidity version evolution and what challenges does it pose?}''
According to official documentation~\cite{solidity080, solidity_github}, the Solidity language has undergone numerous updates, evolving from version \emph{v0.4.1} to \emph{v0.8.23} over the past several years.
This includes 131 ``breaking changes'' during the version evolutions. 
The frequent and ongoing evolution of Solidity has played a pivotal role in driving blockchain technology development. However, these rapid changes have also introduced significant challenges for developers, particularly in terms of knowledge updates, code migration, and ongoing maintenance.
Upon compiling smart contracts across different Solidity versions, approximately 81.68\% (107 out of 131) of these contracts resulted in errors. These errors include parser errors, declaration errors, syntax errors, type errors, JSON errors, and IO errors. Notably, the majority of these errors are compilation errors, accounting for approximately 86.92\% (93 out of 107) of the total. It illustrates that \uline{the rapid evolution of the Solidity language has the potential to induce severe compilation errors, posing substantial challenges for developers (\textbf{Challenge-1}).}

After identifying the challenges posed by Solidity version evolution, we explore effective solutions to address these issues. With the rapid advancement of LLMs~\cite{achiam2023gpt, guo2024deepseek}, these models have demonstrated remarkable performance in various code intelligence tasks, including code generation, program repair, and software testing. However, no existing study has systematically evaluated the capability of LLMs in fixing Solidity compilation errors caused by version migrations.
To bridge this gap, our second research question of the empirical study investigates: ``Can LLMs effectively address the challenges introduced by Solidity version evolution?'' 
To answer this question, we construct a dataset and conduct extensive experiments using two open-source LLMs (i.e., GPT-4o and GPT-3.5-turbo) and two closed-source LLMs (i.e., LLaMA3 and DeepSeek). Inspired by the motivating example in Fig.~\ref{fig:motivating_example}-(b), we design three distinct levels of prompt granularity to generate solutions.
Our findings reveal that different LLMs exhibit varying degrees of effectiveness in resolving Solidity compilation errors, with notable discrepancies across different error types. Crucially, we observe a positive correlation between increased prompt granularity and improved LLM performance in handling these errors.

As demonstrated in the real-world example from Fig.~\ref{fig:motivating_example}-(b), incorporating domain-specific knowledge can significantly enhance LLMs' ability to resolve Solidity compilation errors induced by version migration. However, this introduces a new challenge: \uline{how to acquire accurate expert knowledge for addressing compilation errors caused by version migration (\textbf{Challenge-2})}. This challenge highlights the need for effective strategies to extract and integrate precise domain expertise into the LLM-driven repair process.


In this paper, we propose a novel framework, named \tool, which leverages expert knowledge (e.g., official documentation) and external tools (e.g., static analysis and NLP techniques) to bridge existing gaps and resolve compilation errors when migrating Solidity smart contracts across versions.
\tool consists of three core modules: \textbf{code slicing}, \textbf{knowledge retrieval}, and \textbf{patch generation}, designed to integrate with various LLMs. Specifically, we first construct a domain-specific knowledge base by extracting Solidity version changes from official documentation. Since we focus on the most critical compilation errors in Solidity evolution (\textbf{Challenge-1}), we curate 93 error-inducing changes from a total of 131 documented modifications (as shown in Table~\ref{tab:1}).
To address \textbf{Challenge-2}, we introduce a retrieval mechanism, \retriever, by fine-tuning all-MiniLM-L6-v2\footnote{https://huggingface.co/sentence-transformers/all-MiniLM-L6-v2} using our empirical study dataset. When encountering an uncompileable Solidity code, our approach first applies code slicing to filter out irrelevant code fragments based on the error message. The retrieval module then queries the knowledge base for the most relevant expert knowledge. Finally, using a fine-grained prompt, the patch generation module instructs the LLM to generate a repair patch. The repaired code is recompiled, and if the compilation fails, the process is iteratively repeated until the code is successfully fixed or the maximum iteration limit of five cycles is reached.

We evaluate \tool on both a constructed dataset and a real-world dataset (see Section~\ref{sec:dataset}). Experimental results demonstrate that our approach significantly enhances LLM performance in fixing Solidity compilation errors across different versions. It effectively reduces the performance gap between smaller and larger models, offering a robust and scalable solution for various error types. Notably, on the real-world dataset, our approach achieves a repair accuracy of 96.97\%, outperforming standalone GPT-4o by 24.24\%. Moreover, our ablation study confirms that both code slicing and retrieval mechanisms are critical components, with the retrieval module contributing significantly to performance improvements.

We summarize the contributions of this paper as follows:

\begin{itemize}[wide=0pt]
    \item We are the first to systematically investigate compilation errors caused by Solidity version evolution and evaluate the capability of current LLMs in resolving these errors.
    \item We propose \tool, a framework that systematically integrates expert knowledge, retrieval mechanisms, and LLM-based repair strategies to address Solidity compilation errors induced by version evolution effectively.
    \item We conduct extensive experiments on both a constructed dataset and a real-world dataset, demonstrating that our approach enhances LLM performance in fixing Solidity compilation errors, particularly in narrowing the gap between smaller and larger models. Our ablation study further validates the importance of code slicing and knowledge retrieval.
    \item We publish our code and dataset for other researchers to replicate our study and conduct more future work\footnote{https://github.com/Ylkylkylk/SMCFixer}.   
\end{itemize}

\section{Empirical Study}\label{empirical study}
As is widely recognized, there has been no prior research specifically addressing the evolution of Solidity versions. Therefore, we begin by conducting an empirical study using official online materials to explore the trends and patterns in Solidity version evolution.

\subsection{Research Questions}

This empirical study aims to systematically investigate the current challenges brought by the evolution of the Solidity version and the ability of LLMs to handle these challenges. 
To achieve this goal, we investigate the following two research questions (RQs):

\begin{itemize}[leftmargin=*]
    \item \textbf{RQ1: What is the current state of Solidity version evolution, and what challenges does it pose?}
    We investigate the current trends and challenges associated with the evolution of Solidity versions on the online official materials, aiming to identify key issues that developers and systems face during version migrations.
    \item \textbf{RQ2: Can LLMs help solve the challenges brought by the Solidity version evolutions mentioned above?} We explore the potential of LLMs to mitigate the challenges identified in RQ1, assessing their effectiveness in supporting version migration and code adaptation.
\end{itemize}

\subsection{Solidity Evolution Changes Collection}

Solidity language is well documented online, including specifications of different language versions.
In this case, we parse the Solidity official documentation to obtain the changes of different versions.
This documentation covers the breaking changes from major versions \emph{v0.4.x} to \emph{v0.8.x}, including explicitness requirements, semantic and syntactic changes, new features, interface changes, and so on. 
We employ a web crawling tool~\cite{beautifulsoup4} to extract the relevant content about breaking changes from the official web pages. 
Initially, we extract the content through the \texttt{<section>} tag, which provides the preliminary Solidity evolution changes. Next, we remove all header content marked by \texttt{<h>} tags and use the \texttt{<p>} paragraph tag to obtain detailed descriptions of the changes. 
All HTML tags, except for \texttt{<code>}, are removed, leaving us with the final Solidity Evolution Changes.
As a result, we obtain 131 changes from the official documentation, as shown in Table~\ref{tab:changes_statistics}.

\begin{table}[ht]
\small
\centering
\caption{The Statistics about Solidity Version Evolution}\label{tab:changes_statistics}
\resizebox{0.49\textwidth}{!}{
\begin{tabular}{>{\centering}p{2cm}|c|>{\centering}p{2cm}|c|c|c|c|c|c|c}
\toprule
Version Evolution & \# of Changes & \# of Smart Contracts & \multicolumn{4}{c|}{Compilation Errors} & \multicolumn{2}{c|}{Solc CLI Errors} & No Error\\
\cmidrule(lr){1-1} \cmidrule(lr){4-9} 
\centering X $\rightarrow$ Y & & & Parser & Declaration & Syntax & Type & Json & IO \\
\midrule
\centering 0.4 $\rightarrow$ 0.5 & 57 & 570 & 40 & 40 & 90 & 230 & 60 & 0 & 110\\
\midrule
\centering 0.5 $\rightarrow$ 0.6 & 26 & 260 & 30 & 30 & 0 & 110 & 0 & 20 & 70\\
\midrule
\centering 0.5 $\rightarrow$ 0.7 & \multirow{2}{*}{22} & 220 & 40 & 40 & 10 & 100 & 30 & 0 & 0\\
\cmidrule(lr){1-1} \cmidrule(lr){3-9}
\centering 0.6 $\rightarrow$ 0.7 & & 220 & 40 & 40 & 10 & 100 & 30 & 0 & 0\\
\midrule
\centering 0.5 $\rightarrow$ 0.8 & \multirow{3}{*}{26} & 260 & 20 & 30 & 10 & 110 & 30 & 0 & 60\\
\cmidrule(lr){1-1} \cmidrule(lr){3-9}
\centering 0.6 $\rightarrow$ 0.8 & & 260 & 20 & 30 & 10 & 110 & 30 & 0 & 60\\
\cmidrule(lr){1-1} \cmidrule(lr){3-9}
\centering 0.7 $\rightarrow$ 0.8 & & 260 & 20 & 30 & 10 & 110 & 30 & 0 & 60\\
\midrule
\centering Total &  131& 2050 & 210 & 240 & 140 & 870 & 210 & 20 & 360\\
\bottomrule
\end{tabular}
}
\end{table}

\subsection{Dataset Construction}\label{sec:data_construction}
Solidity is a programming language designed to write smart contracts. Currently, there is no comprehensive dataset dedicated to studying Solidity's version evolution. 
To address this gap, we construct a dataset specifically focused on Solidity evolution changes. 
GPT-4, a state-of-the-art LLM, demonstrates advanced capabilities in processing complex language structures and significantly improves context understanding~\cite{achiam2023gpt4}. 
We leverage GPT-4 to generate instances associated with Solidity version migration issues.

\begin{figure}[htbp]
    \centering
    \includegraphics[width=0.45\textwidth]{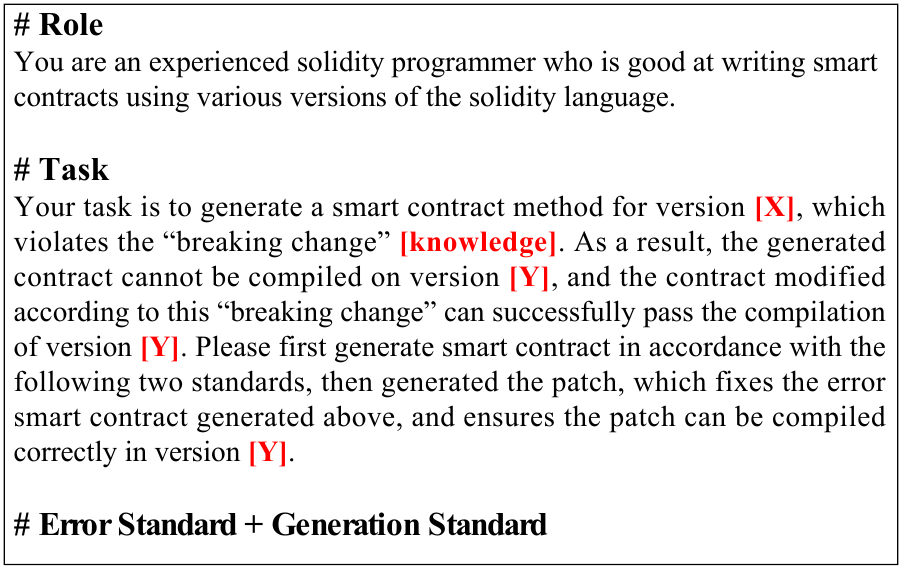}
    \caption{Data Generation Prompt Template}
    \label{fig:datagen}
\end{figure}

The dataset generation task begins by instructing GPT-4 to create a smart contract method for a specific Solidity version (\textcolor{red}{X}) that intentionally violates the documented ''breaking change'' rules when targeting another version (\textcolor{red}{Y}). 
The prompt (see Figure.\ref{fig:datagen}) further requires GPT-4 to generate a patch that resolves the error, ensuring the modified contract successfully compiles on the target version (\textcolor{red}{Y}). 
To ensure the generated instances are high-quality and relevant, the process adheres to two standards: the error standard, which ensures errors align with version update knowledge and limits each example to a single error for clarity, and the generation standard, which promotes diversity through data augmentation techniques such as synonym replacement (SR), random insertion (RI), random swap (RS), and random deletion (RD)~\cite{dong2023boosting}. 
These techniques enhance variability by renaming variables and methods, inserting or swapping functions, and deleting non-essential content while preserving compliance with the Error Standard.

Hence, each instance in the dataset consists of the following components: ``Breaking Changes'', ``Source Version (\textcolor{red}{X})'', ``Target Version (\textcolor{red}{Y})'', ``Error Contract'', and ``Patch''. 
When generating instances, we consider not only the evolution between adjacent versions (e.g., \emph{v0.5.x} to \emph{v0.6.x}), but also the migration across multiple versions (e.g., \emph{v0.5.x} to \emph{v0.8}).
For each evolution-related breaking change, 10 instances were generated. 
In total, this process produced 2050 instances covering 131 breaking changes, as summarized in Table~\ref{tab:changes_statistics}.


After autogeneration, two Solidity experts with two years of experience validated the dataset by cross-checking. The reviewer focuses on the following questions: (1) error patterns against official breaking changes and (2) patch effectiveness in resolving issues. This dual review process guarantees the authenticity of the dataset in capturing real-world migration challenges.

The final dataset, named \dataseta, includes 2050 instances and serves as the foundation for an empirical study addressing two research questions: the challenges posed by Solidity's language evolution (RQ1) and the capability of LLMs, such as GPT-4o, to address these challenges (RQ2). By providing a diverse and high-quality dataset, this work offers valuable insights into managing Solidity version evolution.

\subsection{Challenges of Solidity Version Evolutions (RQ1)}
\subsubsection{Strategy}

To analyze the current state and challenges associated with Solidity version evolution, we adopt the following strategy:
(1) First, we systematically collect and catalog all documented ''breaking changes'' from the official Solidity website~\cite{soliditybreakingchanges2023}. 
This step establishes a comprehensive understanding of the scope and nature of changes introduced across different versions.
(2) Next, we leverage LLMs to automatically generate Solidity smart contracts that incorporate the identified breaking changes (see Section \ref{sec:data_construction} for details). This enables us to simulate real-world scenarios where developers encounter issues due to version migrations.
(3) Finally, we compile these generated smart contracts using Remix~\cite{remixethereum}, an online Solidity compiler, to evaluate the practical challenges posed by version evolution. This process allows us to identify specific issues and inconsistencies arising from the breaking changes, providing insight into the difficulties developers face when migrating between Solidity versions.

\subsubsection{Result}
\begin{figure}[htbp]
    \centering
    \includegraphics[width=0.48\textwidth]{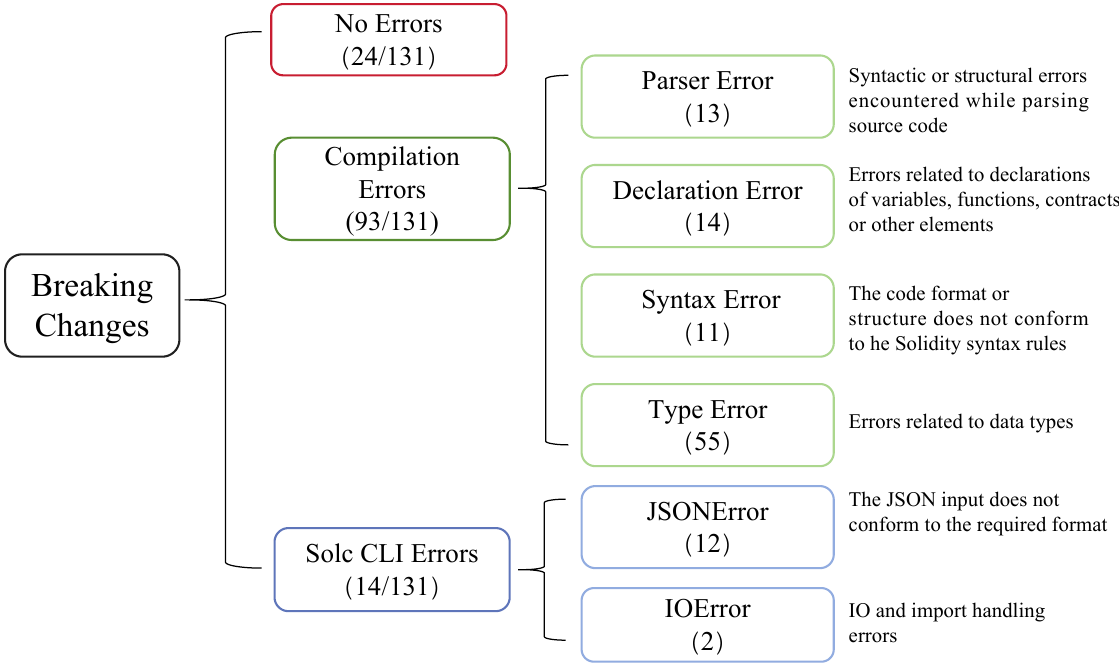}
    \caption{Solidity Version Changes Categorization}
    \label{fig:error_types}
\end{figure}

According to the official Solidity documentation, the language has undergone 84 significant times of evolution over the past several years, with its versions progressing from \emph{v0.4.1} to \emph{v0.8.26}. Table~\ref{tab:changes_statistics} illustrates the number of changes documented, totaling 131 major updates. 
The frequent evolution and numerous major changes across different versions underscore the dynamic development of the Solidity language, highlighting its continuous adaptation and improvement over time. Figure~\ref{fig:error_types} shows that changes that happened during version migration can lead to two types of errors, i.e., compilation errors and Solc CLI errors.

Specifically, compilation errors are the most serious errors caused by Solidity version evolution, accounting for 71\%.
It refers to a situation where the compiler cannot successfully transform the source code into executable machine code due to errors in the code itself.
Specifically, four types of compilation errors happen during the Solidity version evolution:

\begin{itemize}[leftmargin=*]
    \item \textbf{Parser error } refers to syntax or structural errors encountered during the parsing of smart contracts. These errors can include incorrect function calls, inappropriate data type uses, and similar issues. 
    Among the 93 compilation error-related changes, we collect 13 of them that will cause parser errors.
    \item \textbf{Declaration error} pertains to errors related to the declaration of variables, functions, contracts, or other elements, including issues like duplicate declarations, type mismatches, or naming conflicts between variables and functions. 
    In the analysis of 93 compilation error-related changes, 14 cases were identified as declaration error-related.
    \item \textbf{Syntax error} refers to the errors related to the format or structure of the code, which does not adhere to Solidity's syntax rules. These errors include mismatched parentheses, incorrect keywords, and similar issues. 
    Syntax error-related changes are identified in 11 cases.
\item \textbf{Type error} refers to data type errors, including incorrect type conversions, array and mapping-related errors, and so on. 
Among all compilation error-related changes, type errors account for the largest proportion, with 55 out of 93 changes.
\end{itemize}

Solc CLI errors occur when utilizing the Solc compiler through the command-line interface. These errors can be categorized into two types:
\begin{itemize}[leftmargin=*]
    \item \textbf{JSON Error} refers to errors arising when the provided input does not conform to the required JSON format, including instances where the input is not a valid JSON object, the language is unsupported, or other similar format violations. In the analysis of 14 Solc CLI errors-related changes, 12 cases were identified as JSON error-related.
    \item \textbf{IO error} refers to errors related to input/output operations and issues with the processing of imports. Common causes include the inability to resolve URLs or mismatches in hashes within the supplied source files.
    Upon analyzing 14 changes related to Solc CLI errors, it was found that 2 of them were linked to IO errors.
\end{itemize}

\begin{tcolorbox}
\textbf{Answer to RQ1:} 
Solidity's rapid evolution (84 versions, 131 major changes) poses challenges during smart contract migration. Version gaps often trigger compilation errors (71\% of changes introduce such issues), underscoring developers' struggles to keep pace with language updates.
\end{tcolorbox}

\subsection{Capability of LLMs in Handling Challenges (RQ2) }
\subsubsection{Strategy}
According to RQ1's findings, compilation errors are the most serious problem caused by the frequent evolution of the Solidity language.
In RQ2, we would like to investigate whether LLMs can fix such compilation errors.
It is known to all that prompt engineering is widely considered a key factor in eliciting high-quality solutions from LLMs~\cite{arvidsson2023prompt}. 
Inspired by previous work~\cite{ren2023misuse}, we also employ three distinct granularities of prompt design strategies (i.e., general, coarse-grained, and fine-grained prompting) to investigate the capability of LLMs in dealing with compilation errors by the evolution of the Solidity language. 
Figure~\ref{fig:motivating_example}-b shows how we adopt the three granularities of prompts to handle a compilation error raised by developers in Figure~\ref{fig:motivating_example}-a.

Specifically, three types of prompts are designed as follows:

\begin{itemize}[leftmargin=*]
   \item \textbf{General prompting} refers to prompting the model to fix compilation errors by providing only the buggy smart contract code without any additional context or information.
    \item \textbf{Coarse-grained prompting} enhances the prompt by including the error messages reported by the compiler. This provides details about the cause and location of the error, helping the model identify and address the issue more accurately.
    \item \textbf{Fine-grained prompting} further enriches the prompt by incorporating specific version changes and relevant domain knowledge. This enables the LLMs to better understand the root cause of the error and make more precise and informed modifications.
\end{itemize}


We test four LLMs (GPT-4o, GPT-3.5-turbo, LLLaMA3-8B, and Deepseek-7B) to evaluate their pass rates across three prompting strategies for fixing Solidity version-related compilation errors. Pass rate measures the percentage of buggy contracts successfully fixed and compiled without errors after applying these strategies.

\subsubsection{Result}
\begin{table}[htbp]
\centering
\caption{LLM's Pass Rate in Fixing Compilation Errors Under Various Prompting Strategies}
\label{tab:1-1}
\resizebox{0.49\textwidth}{!}{
\begin{tabular}{@{}c|c|c|c|c@{}}
\toprule
LLM & Error Type & General Prompting & Coarse-grained Prompting & Fine-grained Prompting \\ \midrule
GPT-4o & Parser & 58.10\% & 79.05\% & 95.71\% \\
       & Declaration & 56.67\% & 82.50\% & 95.00\% \\
       & Syntax & 53.57\% & 80.71\% & 95.71\% \\
       & Type & 51.38\% & 77.93\% & 92.30\% \\
\midrule
\multicolumn{2}{c|}{\textit{Average}}    & 54.93\% & 80.05\%  & \textbf{94.68\%} \\
\midrule
GPT-3.5-turbo & Parser & 43.81\% & 69.52\% & 89.05\% \\
              & Declaration & 44.58\% & 68.33\% & 90.83\% \\
              & Syntax & 40.71\% & 70.00\% & 91.43\% \\
              & Type & 39.08\% & 69.31\% & 87.01\% \\
\midrule
\multicolumn{2}{c|}{\textit{Average}}    & 42.05\% & 69.29\%  & \textbf{89.58\%} \\
\midrule
LLaMA3-8B & Parser & 44.29\% & 62.86\% & 79.05\% \\
          & Declaration & 51.25\% & 63.75\% & 77.08\% \\
          & Syntax & 49.29\% & 63.57\% & 79.29\% \\
          & Type & 45.86\% & 61.15\% & 75.75\% \\
\midrule
\multicolumn{2}{c|}{\textit{Average}}    & 47.67\% & 62.83\%  & \textbf{77.79\%}\\
\midrule
Deepseek-7B & Parser & 54.29\% & 66.67\% & 90.95\% \\
            & Declaration & 54.17\% & 71.25\% & 91.25\% \\
            & Syntax & 54.29\% & 67.86\% & 93.57\% \\
            & Type & 53.91\% & 67.01\% & 90.11\% \\ 
\midrule
\multicolumn{2}{c|}{\textit{Average}}    & 54.16\% & 68.20\%  & \textbf{91.47\%}  \\
\bottomrule
\end{tabular}
}
\end{table}

Table~\ref{tab:1-1} provides a comprehensive overview of the pass rates for four LLMs in fixing different types of compilation errors in Solidity smart contracts. These results highlight the performance of the models across three distinct prompting strategies: general prompting, coarse-grained prompting, and fine-grained prompting.
\begin{itemize}
    \item \textbf{Performance of LLM models} 
    The models show notable differences across error types and prompting strategies. GPT-4o and Deepseek-7B lead with pass rates of 94.68\% and 91.47\% under fine-grained prompting, respectively, indicating strong generalization in handling Solidity compilation errors. In comparison, GPT-3.5-turbo scores 89.58\%, and LLaMA3-8B trails at 77.79\%, particularly struggling with complex errors.
\item \textbf{Performance of prompt granularity} 
Our analysis shows a strong correlation between prompt granularity and pass rate. On average, models see a 38.69\% improvement when shifting from general to fine-grained prompting, with GPT-3.5-turbo gaining 47.53\%. GPT-4o achieves 95.71\% on syntax errors under fine-grained prompting, up from 53.57\%, while Deepseek-7B improves by 39.4\%. These results highlight the critical role of detailed prompts in enhancing LLM performance for error resolution.
\item \textbf{Performance of error types} 
Our analysis reveals significant accuracy variations across compilation error types. While declaration and syntax errors achieve high resolution rates (e.g., GPT-4o: 95.71\% on syntax errors), type errors remain the most challenging (e.g., GPT-4o: 92.30\%). Smaller models like LLaMA3-8B struggle notably, dropping to 75.75\% on type errors, reflecting LLMs' broader limitations in complex reasoning tasks. These results underscore that type errors demand deeper understanding of Solidity’s type system, a persistent hurdle for LLMs despite prompting optimizations.
\end{itemize}

\begin{tcolorbox}
\textbf{Answer to RQ2:} 
LLMs exhibit varying capabilities in resolving Solidity compilation errors, with performance differing significantly across error types. Increasing prompt granularity directly enhances their effectiveness in addressing these issues.
\end{tcolorbox}

\section{Approach}

\begin{figure}[htbp]
    \centering
    \includegraphics[width=0.49\textwidth]{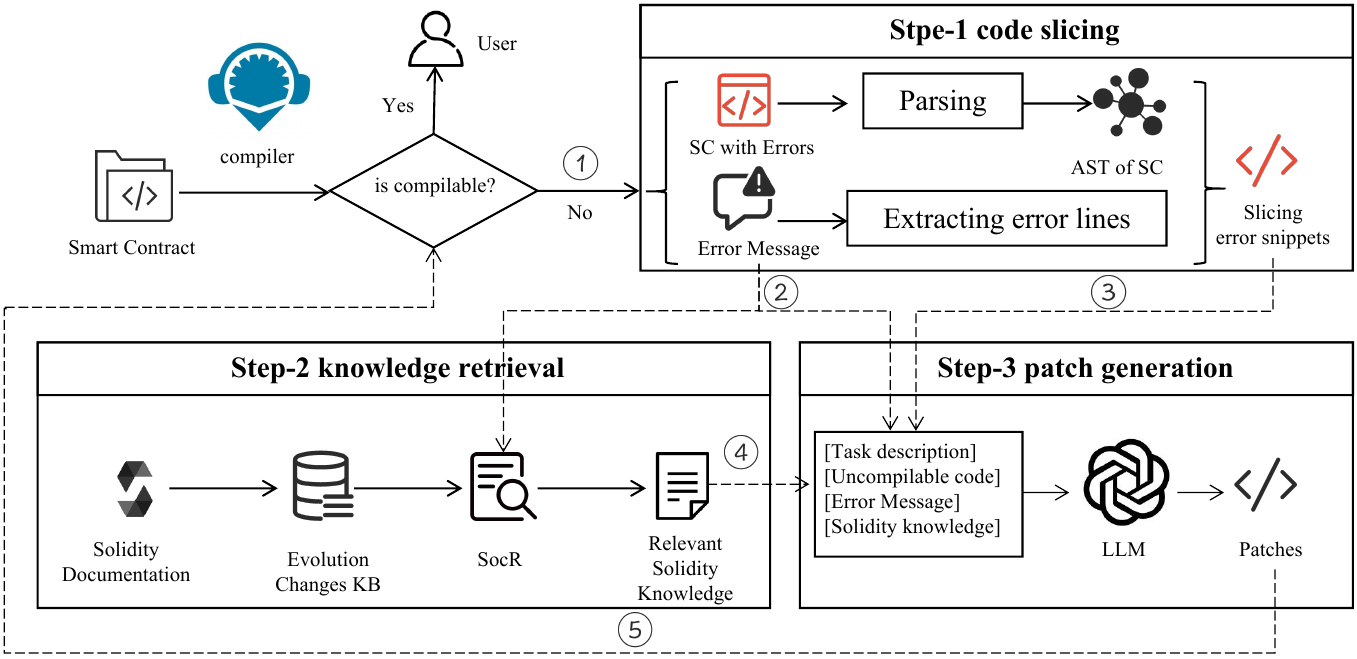}
    \caption{Overall Framework of \tool}
    \label{fig:frame}
\end{figure}

Figure~\ref{fig:frame} illustrates the overall architecture of \tool, comprising three key modules: code slicing, knowledge retrieval, and patch generation. 
Given a Solidity code file, we first use Remix~\cite{remixethereum} to attempt compilation. If the file fails to compile, our approach automatically generates patches by leveraging the compiler’s error messages and relevant expert knowledge.
The subsections provide a detailed explanation of the three core modules of \tool.

\subsection{Code Slicing}
\label{subsec:code slicing}

\begin{figure*}[htbp]
    \centering
    \includegraphics[width=1\textwidth]{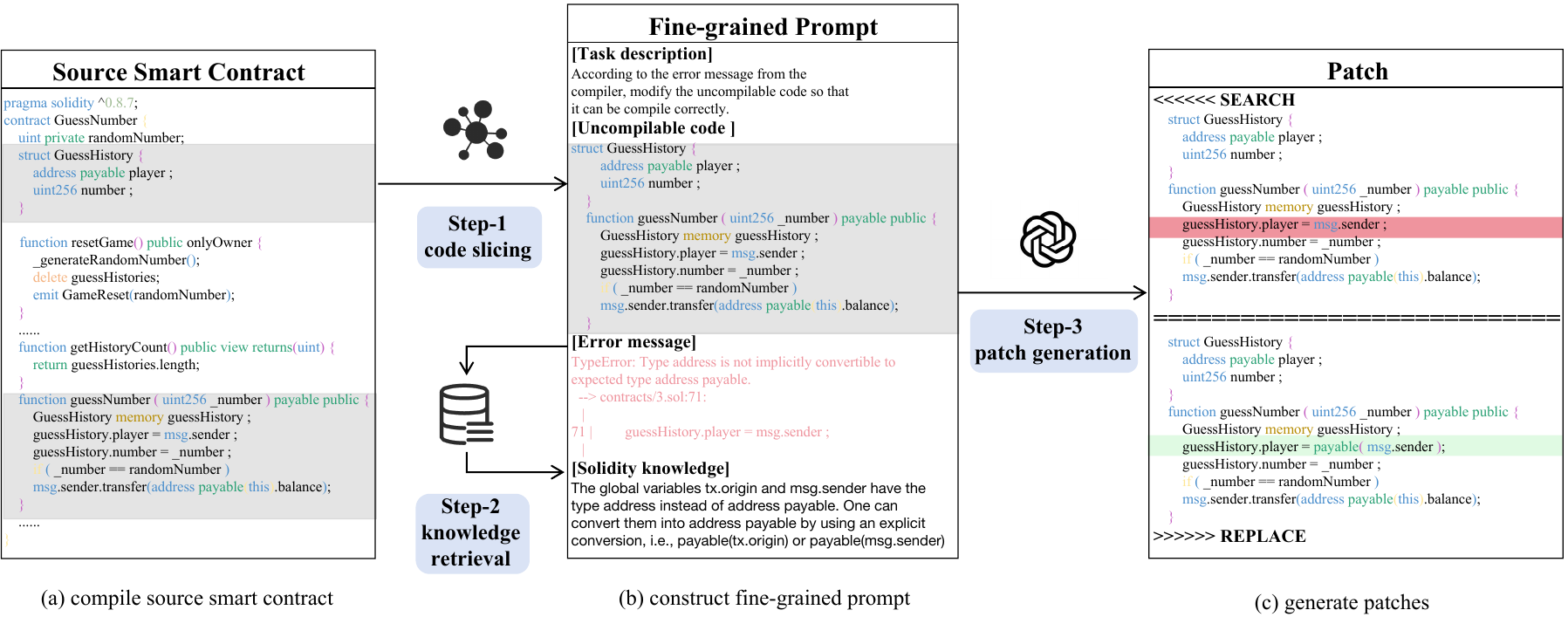}
    \caption{Running Example of \tool}
    \label{fig:running_example}
\end{figure*}

Solidity files define smart contract logic but often face compilation errors during version migration. While LLMs excel at code analysis~\cite{ren2023misuse,yang2024swe,chen2024coder}, feeding entire files to LLMs is inefficient due to context window limits and input noise. Compiler error messages also lack sufficient context to resolve all issues.
For example, in Figure~\ref{fig:running_example}-(b), the error message shows that a type error at line 71 (assigning the address of the current contract caller to the \texttt{player} field in the \texttt{guessHistory} structure) originates from the type declaration at line 5. Fixing such errors requires isolating both compiler-flagged lines and semantically related code snippets (e.g., variable definitions) to address root causes effectively.




We propose a code-slicing method to isolate relevant code snippets for LLM prompts. Key steps: (1) AST parsing: Convert Solidity code into an Abstract Syntax Tree (AST) using \texttt{python-solidity-parser}~\cite{solidity_parser}. This AST serves as a structured representation of the contract, where each node corresponds to a specific syntactic element such as variable declarations, function definitions, expressions, or control flow statements.
(2) Error line extraction: Next, we analyze the compiler error messages (e.g., from solc) to extract line numbers associated with the reported error. We iterate over the leaf nodes of the AST and match their source location metadata (e.g., start line, end line, and column) against the error line. Once a match is found, we retain nodes whose content matches predefined error keywords. To capture contextual dependencies, we expand the selection to include one layer of structurally connected nodes. These typically involve the parent statement node and semantically relevant siblings such as variable initializations, usage references, or surrounding control blocks.
(3) Slicing error snippets: Finally, based on the filtered and expanded AST nodes, we extract corresponding code snippets from the original source file. Each snippet includes not only the immediate error-inducing statement but also adjacent statements or declarations that are semantically coupled through the AST.
For example, as shown in Figure~\ref{fig:running_example}, use the element \texttt{player} in the error message to obtain code snippets from the source smart contract. These code snippets are then used to construct prompts for the LLM.


\subsection{Knowledge Retrieval}
\label{subsec:knowledge retrieval}

Building on RQ2's findings about the necessity of domain knowledge for Solidity version migration, we detail the construction of our expert knowledge base and the design of our retriever to enable efficient, accurate knowledge retrieval.


\subsubsection{Knowledge database construction}

Solidity has evolved through multiple versions (e.g., \emph{v0.4.x} to \emph{v0.8.x}), each introducing breaking changes, syntax updates, and security improvements. LLMs struggle to track these frequent updates due to training data limitations~\cite{raiaan2024review}. To mitigate this limitation, we build a retrieval-augmented generation (RAG) framework~\cite{gao2023retrieval, ren2023misuse} with a domain-specific knowledge base, enabling LLMs to dynamically access version migration rules and ensure accurate fixes.


As described in RQ1, we curated 93 major Solidity updates linked to compilation errors (from 131 crawled from official documents) for our knowledge base (see Section~\ref{empirical study} for details). Raw HTML data was cleaned by removing extraneous tags while preserving critical \texttt{<code>} blocks, which enhance code context understanding. The structured knowledge base integrates these version-specific changes to support accurate retrieval and patch generation for LLMs.


\subsubsection{Retriever design}

Generic RAG frameworks struggle with domain-specific tasks like code repair due to inefficient knowledge structuring and irrelevant retrieval, as noted in prior studies~\cite{zhao2024retrieval, barnett2024seven}. We address this by developing a domain-optimized retriever and fine-tuning its embeddings on our Solidity knowledge base to enable precise, context-aware retrieval of version-specific fixes.


Our retriever is a variant of \textit{all-MiniLM-L6-v2}~\cite{all-MiniLM-L6-v2}, a 6-layer transformer model. 
To enhance its sensitivity to code, we incorporated a multi-head self-attention module, which assigns higher weights to code in the dataset. 
We replaced the standard \texttt{ReLU} activation function with the \texttt{GELU} activation function, which provides smoother non-linear transformations and performs better in natural language processing tasks~\cite{radford2018gpt}.

The model was fine-tuned on our custom dataset (\dataseta, see Section~\ref{empirical study}) using the \texttt{CosineSimilarityLoss} function, applied to triples of structured data: \texttt{<query>}, \texttt{<answer>}, and \texttt{<label>}.


\begin{itemize}
    \item \texttt{<query>} represents compiler-generated error messages that were preprocessed to remove redundant or extraneous text (e.g., phrases such as ``Error --\textgreater{}'').
    \item \texttt{<answer>} corresponds to examples retrieved from the knowledge base, which serve as potential solutions or explanations for the error messages.
    \item \texttt{<label>} is a binary indicator (0 or 1) that specifies whether the error message is relevant to the knowledge base (1) or not (0).
\end{itemize}

The dataset (\dataseta) was split into 80\% training and 20\% testing to balance model training and generalization evaluation. Fine-tuning employed the \texttt{AdamW} optimizer (learning rate: 2e-5, batch size: 16) for 100 epochs, with early stopping triggered if testing accuracy stalled for 5 epochs. This configuration ensured stable convergence while optimizing computational efficiency.



\subsection{Patch Generation}

This study uses LLMs to fix Solidity compilation errors by generating code patches, leveraging their semantic understanding and code repair capabilities. Effective prompt design is critical, providing enough context for accurate fixes while avoiding excessive input noise that degrades performance.


The fine-grained prompting strategy (see Figure~\ref{fig:frame} Step-3) structures inputs into four blocks: (1) \texttt{[Task description]}: defines the role and objectives of the LLM. (2) \texttt{[Uncompilable code]}: includes code snippets isolated via code slicing. (3) \texttt{[Error meaasge]}: Displays compiler-reported error details. (4) \texttt{[Solidity knowledge]}: Provides version-specific rules retrieved from the knowledge base in~\ref{subsec:knowledge retrieval}. Together, these blocks help the LLM to generate precise fixes for compilation errors.

We designed an output format based on *search/replace* edits to automate the patch application process. The format includes:
\begin{itemize}
    \item The start of the search block: \texttt{<<<<<<\ SEARCH}.
    \item A contiguous chunk of lines to search in the source code.
    \item The dividing line: \texttt{=======}.
    \item The lines to replace in the source code.
    \item The end of the replace block: \texttt{>>>>>>\ REPLACE}.
\end{itemize}

Patches are applied via git apply, followed by recompilation. If errors persist, the system retries fixing iteratively until compilation succeeds or the maximum attempt limit is reached.


\section{Experimental Setup}
\subsection{Research Question}
We propose the following three research questions (RQ) to validate the effectiveness of \tool:
\begin{itemize}[leftmargin=*]
\item \textbf{RQ3: How does our approach perform in fixing errors?} We compare our approach with open-source and closed-source LLMs to verify the overall performance of \tool fixing compile errors.
\item \textbf{RQ4: How do our selected design choices perform?} We first perform ablation studies to validate the effectiveness of the module, then evaluate the performance of the retrieval approach.
\item \textbf{RQ5: How does our approach perform in handling real-world completion errors?} We test the effectiveness of our approach using GitHub/Stack Overflow smart contracts.
\end{itemize}

\subsection{Baseline}

We select four distinct LLMs to evaluate our approach's performance, including two close-source LLMs (i.e., GPT-4o~\cite{hurst2024gpt4o} and GPT-3.5-trubo~\cite{ye2023gpt3}) and two open-source LLMs (i.e., LLaMA3-8B~\cite{dubey2024llama} and Deepseek-7B~\cite{guo2024deepseek}).
To assess the overall effectiveness of \tool, we employ these four LLMs (i.e., GPT-4o, GPT-3.5-turbo, LLaMA3-8B, and Deepseek-7B) as end-to-end LLM-based baselines.

Furthermore, we create several variants on Module-1 (code slicing) and Module-2 (knowledge retrieval) to demonstrate the impact of different design choices within \tool.
The following variants are examined:

\begin{itemize}
    \item \textbf{\tool{\_w/o\_slicing}} This variant removes the code slicing module, where the original Solidity code is treated as \texttt{[uncompilable code]}. By comparing \tool to \tool{\_w/o\_slicing}, we evaluate the effectiveness of code slicing.
    \item \textbf{\tool{\_w/o\_retriever}} This variant disables retrieval, leaving prompts knowledge-deficient during generation. Comparing \tool with its retrieval-less version assesses how specialized knowledge addresses Solidity compilation errors.
\end{itemize}


We benchmark our retriever against the retrieval baselines to evaluate component performance and knowledge quality.

\begin{itemize}
    \item \textbf{BM25+DPR}~\cite{lu2022reacc} BM25 uses traditional TF-IDF methods for precise retrieval, while Dense Passage Retrieval (DPR) employs deep learning to capture semantic meaning. Their combination effectively meets both exact and semantic query needs.
    \item \textbf{Nv-embed-v1}~\cite{lee2024nvembed} A transformer-based model generates dense embeddings for retrieval/similarity tasks via contrastive learning, enhancing related pairs' similarity while suppressing unrelated ones.
    \item \textbf{Rerank-qa-mistral}~\cite{nvidia2024rerank} An optimized model predicts relevance probabilities for question-answering in passages. Ranking models in text retrieval systems aim to enhance accuracy.
    \item \textbf{All-MiniLM-L6-v2} It is a lightweight pre-trained model that is a variant of the MiniLM~\cite{wang2020minilm} series, designed to provide efficient performance with relatively low computational resource consumption for NLP tasks.
\end{itemize}

\subsection{Dataset}\label{sec:dataset}

\subsubsection{Experimental Dataset}

We employ two datasets to evaluate the effectiveness of our approach: \datasetb and \datasetc.

\begin{itemize}
    \item \textbf{\datasetb} A dataset is constructed by generating 10 contracts per evolutionary change causing compilation errors, totaling 1,460 contracts based on Solidity language updates.
    \item \textbf{\datasetc} A real-world dataset of 33 contracts from GitHub/OpenZeppelin/reddit/stack exchange, all failing due to Solidity version syntax or semantic changes.
\end{itemize}


\subsection{Metric}

We use the following metrics for evaluation:

\begin{itemize}[leftmargin=*]

\item \textbf{Pass Rate}: The compilation pass rate measures the percentage of error-free compilations, indicating how effectively a method resolves Solidity version-induced errors. The formula is: $Pass\ Rate = \frac{Successful\_compiled\_contracts}{All contracts}\times 100$


\item \textbf{Acc@k}: It represents the top-K retrieval accuracy of the knowledge base, which measures the proportion of relevant items retrieved within the top $k$ results. 
The \textit{acc@k} is: $Acc@k = \frac{Retrieved\_item\_number}{k}\times 100$.

\item \textbf{BLEU-4}: BLEU-4 evaluates how similar a generated patch is to the correct one by measuring the 4-gram overlap between them.


\item \textbf{Edit Similarity (ES)}: ES measures similarity between generated and groundtruth patches based on minimum edits. Scores closer to 1 indicate higher similarity.


\end{itemize}


\section{Evaluation}
\subsection{Overall Performance (RQ3)}
\subsubsection{Evaluation Strategy}

We test \tool on \datasetb (1,460 Solidity contracts with version-related compilation errors) using four LLMs: GPT-4o, GPT-3.5-turbo, LLaMA3-8B, and Deepseek-7B. We evaluate performance via compilation pass rate and alignment metrics (BLEU-4/ES) between generated patches and ground truth.



\subsubsection{Result}

\begin{figure}[htbp]
    \centering
    \includegraphics[width=0.49\textwidth]{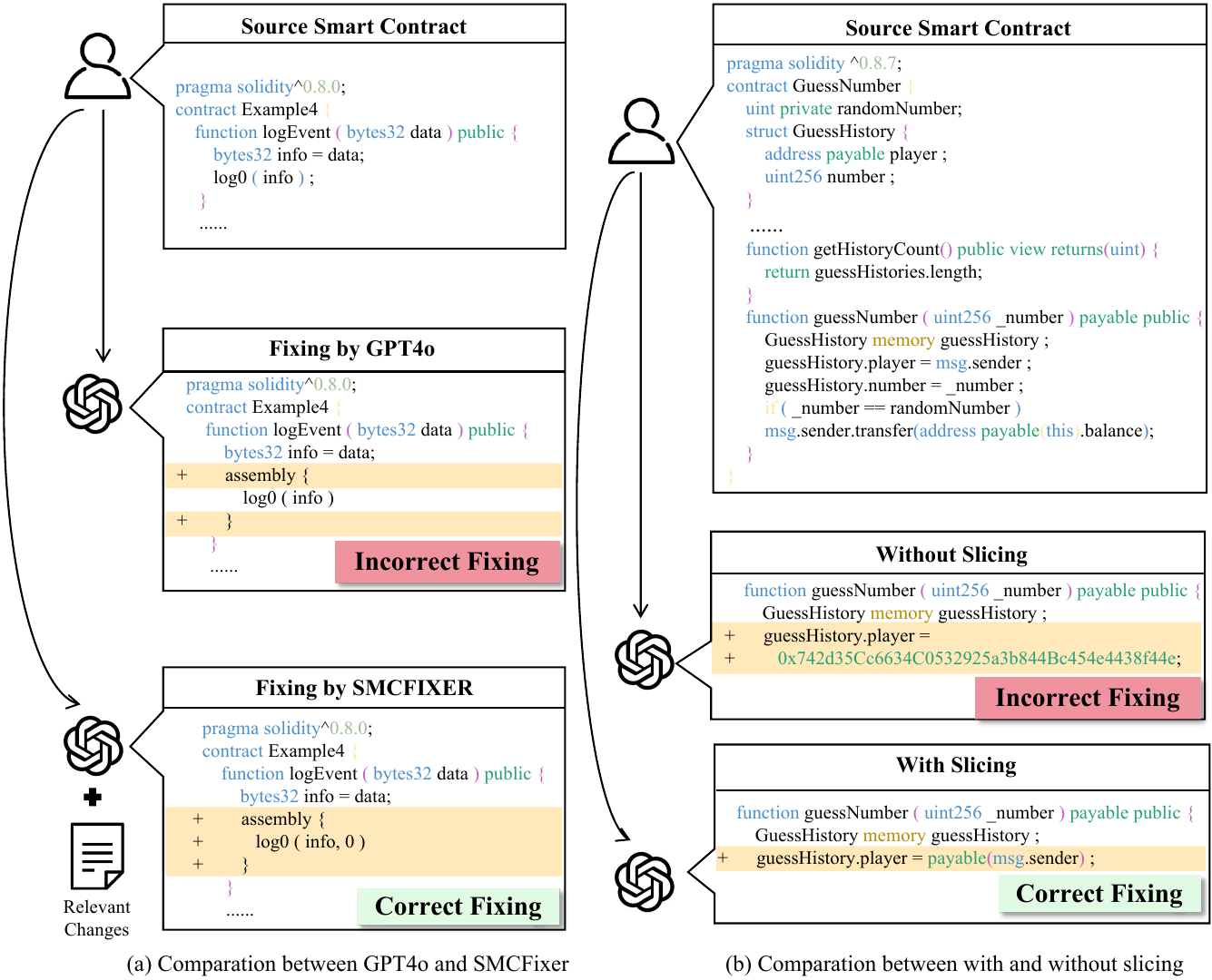}
    \caption{Fixing Example of RQ3 and RQ4}
    \label{fig:rq3-4}
\end{figure}

Table~\ref{tab:overall} demonstrates \tool's superiority in resolving Solidity errors compared to standalone LLMs. Vertically, our approach achieves a 95.20\% average pass rate with GPT-4o (+15.89\% improvement) and elevates smaller models like LLaMA3-8B from 64.06\% to 88.43\% (+24.37\%), bridging performance gaps across model scales. Its model-agnostic design enables adaptability to diverse computational resources. A case study (Figure~\ref{fig:rq3-4}-a) highlights \tool's precision in addressing version-specific changes, such as adapting the deprecated \texttt{log0} function in Solidity \emph{v0.8.x}, where standalone GPT-4o failed to adjust argument requirements.

Horizontally, our approach enhances flexibility by addressing broader error types and narrowing performance disparities between models. Smaller models like LLaMA3-8B and Deepseek-7B see significant gains (e.g., +24.37\% pass rate for LLaMA3-8B), reducing capability gaps with larger models like GPT-4o. This underscores the ability to mitigate the inherent limitations of smaller LLMs, democratizing access to high-quality error resolution regardless of model size.

Our approach excels in handling Solidity language evolution errors. It improves pass rates for declaration errors (97.92\% vs. 82.92\% with standalone GPT-4o) and type errors (92.99\% vs. 76.78\%), addressing syntax and typechecking changes across versions. These results emphasize our approach tailored optimizations for Solidity’s evolving ecosystem, ensuring robust compatibility with version-specific language rules.

\begin{table}[ht]
\centering
\caption{Overall Performance of \tool in Fixing Solidity Errors}
\label{tab:overall}
\small
\resizebox{0.49\textwidth}{!}{
\begin{tabular}{@{}c|c|c|c|c|c@{}}
\toprule
\textbf{Approach}               & \textbf{LLM}          & \textbf{Error Type} & \textbf{Pass Rate}  & \textbf{BLEU-4} & \textbf{ES}   \\ \midrule
\multirow{20}{*}{End-to-end LLMs}
                            & \multirow{5}{*}{GPT-4o} & Parser        & 79.52\% & 93.45\%  & 91.77\% \\
                            &                        & Declaration   & 82.92\% & 97.68\%  & 96.19\% \\
                            &                        & Syntax        & 80.00\% & 96.28\%  & 98.30\% \\
                            &                        & Type          & 76.78\% & 92.15\%  & 93.96\% \\
                            \cmidrule(l){3-6}
                            &                        &{\textit{Average}}    & 79.31\% & 94.89\%  & 94.88\% \\ \cmidrule(l){2-6}
                            & \multirow{5}{*}{GPT-3.5-turbo} & Parser    & 67.62\% & 87.75\%  & 91.41\% \\
                            &                        & Declaration   & 71.67\% & 90.42\%  & 95.53\% \\
                            &                        & Syntax        & 70.00\% & 93.11\%  & 92.34\% \\
                            &                        & Type          & 65.75\% & 89.08\%  & 89.71\% \\
                            \cmidrule(l){3-6}
                            &                        &{\textit{Average}}    & 68.76\% & 90.09\%  & 92.25\% \\ \cmidrule(l){2-6}
                            & \multirow{5}{*}{LLaMA3-8B} & Parser      & 62.38\% & 84.93\%  & 95.01\% \\
                            &                        & Declaration   & 67.08\% & 86.38\%  & 99.08\% \\
                            &                        & Syntax        & 62.86\% & 85.14\%  & 93.86\% \\
                            &                        & Type          & 63.91\% & 86.18\%  & 93.43\% \\
                            \cmidrule(l){3-6}
                            &                        &{\textit{Average}}    & 64.06\% & 85.66\%  & 96.22\% \\ \cmidrule(l){2-6}
                            & \multirow{5}{*}{Deepseek-7B} & Parser     & 66.19\% & 93.39\%  & 95.17\% \\
                            &                        & Declaration   & 71.25\% & 95.51\%  & 99.04\% \\
                            &                        & Syntax        & 68.57\% & 93.90\%  & 98.75\% \\
                            &                        & Type          & 65.52\% & 90.04\%  & 94.88\% \\
                            \cmidrule(l){3-6}
                            &                        &{\textit{Average}}    & 67.88\% & 93.21\%  & 96.96\% \\ \midrule
\multirow{20}{*}{SMCFIXER}         
                            & \multirow{5}{*}{GPT-4o} & Parser  & 94.76\% & 96.85\%  & 96.82\% \\
                            &                        & Declaration   & 97.92\% & 99.82\%  & 99.85\% \\
                            &                        & Syntax        & 97.14\% & 98.36\%  & 99.95\% \\
                            &                        & Type          & 92.99\% & 96.61\%  & 97.28\% \\
                            \cmidrule(l){3-6}
                            &                        &{\textit{Average}}    & \textbf{95.20\%($\uparrow$15.89\%)} & \textbf{98.91\%($\uparrow$4.02\%)}  & \textbf{99.19\%($\uparrow$4.31\%)}\\ \cmidrule(l){2-6}
                            & \multirow{5}{*}{GPT-3.5-turbo} & Parser & 88.57\% & 97.77\%  & 93.16\% \\
                            &                        & Declaration   & 92.50\% & 98.45\%  & 97.86\% \\
                            &                        & Syntax       & 89.29\% & 97.04\%  & 97.24\% \\
                            &                        & Type         & 88.28\% & 95.59\%  & 92.94\% \\
                            \cmidrule(l){3-6}
                            &                        &{\textit{Average}}    & \textbf{89.66\%($\uparrow$20.90\%)} & \textbf{97.21\%($\uparrow$7.12\%)}  & \textbf{95.43\%($\uparrow$3.18\%)}\\ \cmidrule(l){2-6}
                            & \multirow{5}{*}{LLaMA3-8B} & Parser & 87.62\% & 92.96\%  & 97.23\% \\
                            &                        & Declaration   & 91.67\% & 96.92\%  & 99.65\% \\
                            &                        & Syntax        & 88.57\% & 94.39\%  & 94.08\% \\
                            &                        & Type          & 85.86\% & 90.73\%  & 95.25\% \\
                            \cmidrule(l){3-6}
                            &                        &{\textit{Average}}    & \textbf{88.43\%($\uparrow$24.37\%)} &\textbf{93.69\%($\uparrow$8.03\%)}  & \textbf{96.27\%($\uparrow$0.05\%)}\\ \cmidrule(l){2-6}
                            & \multirow{5}{*}{Deepseek-7B} & Parser & 91.43\% & 99.34\%  & 96.65\% \\
                            &                        & Declaration   & 95.83\% & 99.53\%  & 99.34\% \\
                            &                        & Syntax       & 94.29\% & 96.86\%  & 96.84\% \\
                            &                        & Type          & 90.69\% & 95.84\%  & 95.25\% \\
                            \cmidrule(l){3-6}
                            &                        &{\textit{Average}}    & \textbf{93.06\%($\uparrow$25.18\%)} & \textbf{97.89\%($\uparrow$4.68\%)}  & \textbf{97.02\%($\uparrow$0.06\%)}\\ \bottomrule
\end{tabular}
}
\end{table}

\begin{tcolorbox}
\textbf{Answer to RQ3:} 
Our method boosts LLMs' ability to fix Solidity errors from version updates, surpassing standalone models and reducing performance disparities between smaller and larger LLMs in diverse errors.
\end{tcolorbox}

\subsection{Performance of Different Designing Choices (RQ4)}
\lstdefinestyle{conferencestyle}{
    backgroundcolor=\color{white},  
    commentstyle=\color{codegray},  
    keywordstyle=\bfseries,         
    numberstyle=\tiny\color{codegray},  
    stringstyle=\color{codepurple},  
    basicstyle=\ttfamily\tiny,  
    breakatwhitespace=false,       
    breaklines=true,               
    captionpos=b,                  
    keepspaces=true,               
    numbers=left,                  
    numbersep=3.5pt,                 
    showspaces=false,              
    showstringspaces=false,        
    showtabs=false,                
    tabsize=2,                     
    frame=single,                  
    frameround=fttt,               
    framexleftmargin=-2.5pt,          
    framexrightmargin=-2pt,         
    columns=flexible               
}
\lstset{style=conferencestyle}

\subsubsection{Evaluation Strategy}

We assess key design choices in our approach framework using \datasetb: (1) measuring retrieval accuracy (Acc@k) of the \retriever module; (2) testing variants \tool{\_w/o\_slicing} (no code slicing) and \tool{\_w/o\_retrieval} (no retrieval), both built on GPT-4o, to evaluate module contributions; (3) analyzing the multi-iteration strategy by varying loop limits (1–10 iterations).



\subsubsection{Result}


\noindent (1)~\textbf{Performance of retrieval approach:} Table~\ref{tab:4} shows our retrieval module (\retriever) outperforms baselines across ranking thresholds (81.34\% at Acc@1, 91.75\% at Acc@3, and 94.5\% at Acc@5), attributed to fine-tuning on version migration cases and enhanced sensitivity to Solidity code in compiler error messages (see Figure~\ref{fig:error message}). For example, when handling code snippets like \texttt{msg.sender.transfer(...)}, \retriever accurately retrieves version-specific rules (e.g., address payable conversions in \emph{v0.8.x}), while semantic/textual similarity-based retrievers fail to prioritize code-centric context. This ensures precise, relevant knowledge retrieval for error fixes.

\begin{table}[htbp]
\centering
\caption{Accuracy of Various Retrieval Approaches.}
\resizebox{0.49\textwidth}{!}{
\begin{tabular}{l|c|c|c}
\toprule
\textbf{Approach} & \textbf{Acc@1} & \textbf{Acc@3} & \textbf{Acc@5}\\
\midrule
BM25+DPR & 36.47\%($\downarrow$44.87\%) & 50.38\%($\downarrow$41.37\%) & 62.62\%($\downarrow$31.88\%) \\
nv-embed-v1 & 65.38\%($\downarrow$15.96\%) & 89.23\%($\downarrow$2.52\%) & 90.77\%($\downarrow$3.73\%)\\
rerank-qa-mistral & 78.31\%($\downarrow$3.03\%) & 88.89\%($\downarrow$2.86\%) & 90.11\%($\downarrow$4.39\%)\\
all-MiniLM-L6-v2  & 56.12\%($\downarrow$25.22\%) & 76.54\%($\downarrow$15.21\%) & 83.73\%($\downarrow$10.77\%)\\
\midrule
Our \retriever  & \textbf{81.34\%} & \textbf{91.75\%} & \textbf{94.50\%}\\
\bottomrule
\end{tabular}
}
\label{tab:4}

\end{table}

\lstset{
    basicstyle=\ttfamily,
    moredelim={[is][\color{red}]{@}{@}} 
}

\begin{figure}[H]
    \centering
\begin{lstlisting}[basicstyle=\footnotesize]
TypeError: "send" and "transfer" are only available for objects of type "address payable", not "address".
  --> contracts/3.sol:27:9:
   |
27 |@         msg.sender.transfer(payable(address(this)).balance);@
   |         ^^^^^^^^^^^^^^^^^^^
\end{lstlisting}
\caption{Example of a Solidity Compiler Error Message.}
\label{fig:error message}
\end{figure}

\noindent (2)~\textbf{Performance of module ablation:}
Table~\ref{tab:ablation} highlights the impact of removing key modules from \tool. Omitting the code slicing module ({\_w/o\_slicing}) marginally reduces performance but critically weakens handling of longer contracts by failing to isolate error-prone code segments (e.g., incorrect \texttt{address payable} assignments in Figure~\ref{fig:rq3-4}-(b)). While less vital for shorter code, slicing enhances robustness in complex scenarios. Removing the retrieval module ({\_w/o\_retrieval}) causes severe drops: pass rate (-14.92\%), BLEU-4 (-3.20\%), and ES (-3.84\%). Without retrieval, LLMs lack context for fixes (e.g., missing \texttt{log0} parameter updates in Figure~\ref{fig:rq3-4}-(a)), underscoring their role in supplying Solidity-specific knowledge. The retrieval mechanism is indispensable for resolving version-specific errors, as shown by its absence leading to incomplete fixes. Code slicing, though less impactful overall, proves essential in reducing noise for larger contracts. Both modules collectively ensure \tool's accuracy and adaptability across diverse contract complexities.

\begin{table}[htbp]
\centering
\caption{Results of the Ablation Study.}\label{tab:ablation}
\resizebox{0.49\textwidth}{!}{
\begin{tabular}{l|c|c|c}
\toprule
\textbf{Approach} & \textbf{Pass Rate} & \textbf{BLEU-4} & \textbf{ES} \\
\midrule
\tool{\_w/o\_slicing}        & 92.40\% ($\downarrow$2.80\%) &  98.01\% ($\downarrow$0.9\%) & 98.69\% ($\downarrow$0.5\%) \\
\tool{\_w/o\_retrieval}      & 80.75\% ($\downarrow$14.93\%) & 95.71\% ($\downarrow$3.2\%) & 95.35\% ($\downarrow$3.84\%) \\
\midrule
Our \tool         & \textbf{95.20\%} & \textbf{98.91\%} & \textbf{99.19\%} \\                  
\bottomrule
\end{tabular}
}
\end{table}

\begin{figure}[htbp]
    \centering
    \includegraphics[width=0.4\textwidth]{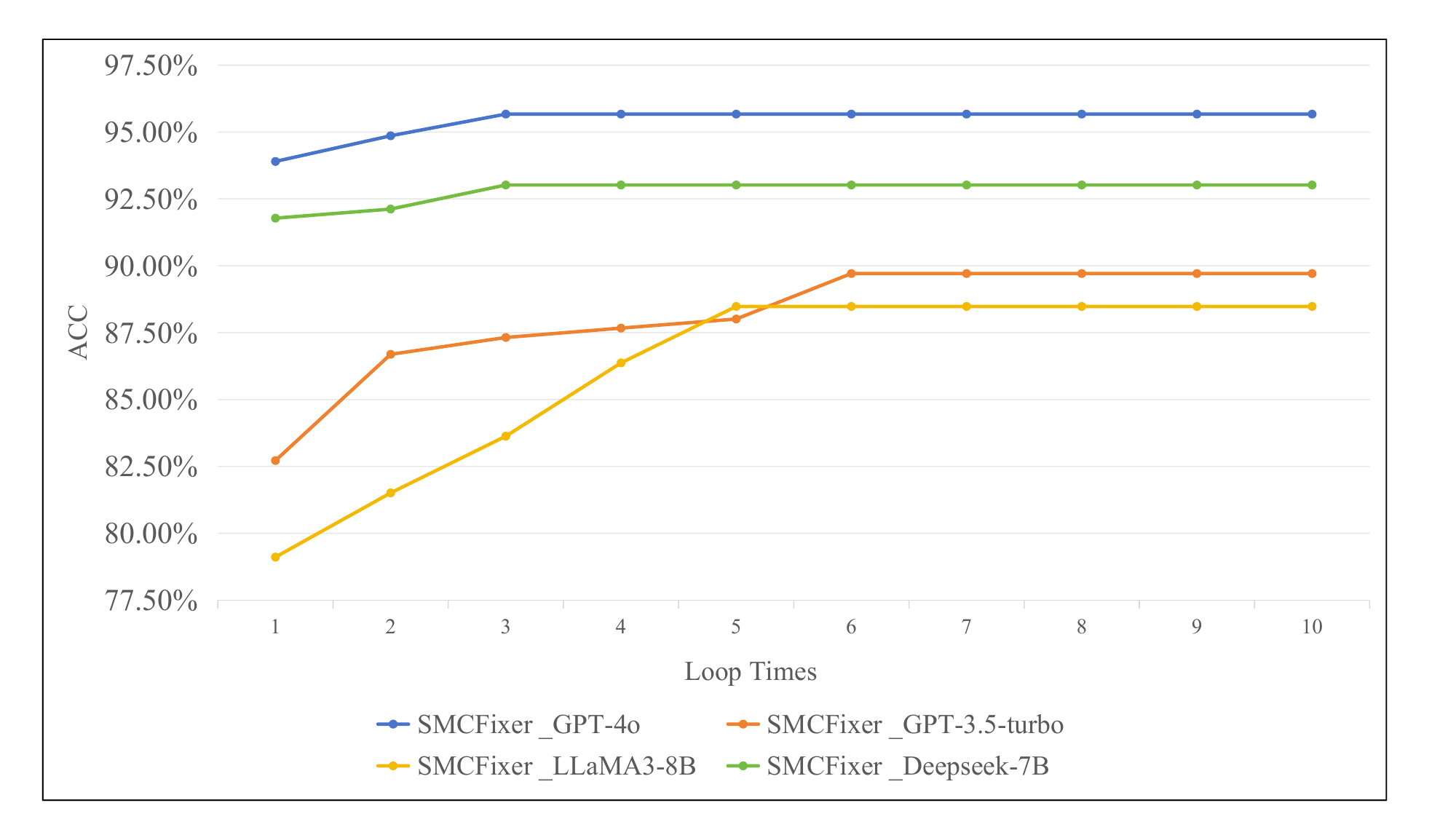} 
    \caption{Compilation Pass Rate in Various Maximum Loops.}
    \label{fig:loop}
\end{figure}

\noindent (3)~\textbf{Performance of iteration strategy:}
Figure~\ref{fig:loop} demonstrates the effect of multi-iteration strategies on performance. GPT-4o and Deepseek-7B achieved stable pass rates (by$>$93.00\%) within 3 iterations, with diminishing returns beyond this point. LLaMA-3-8B and GPT-3.5-turbo required 5–6 iterations to stabilize at 88.48\% and 89.71\%, respectively. These results indicate that a maximum of 5 iterations optimizes efficiency and effectiveness, balancing performance gains for smaller models while avoiding redundant computations as improvements plateau afterward.

\begin{tcolorbox}
\textbf{Answer to RQ4:} 
Our approach relies on two key components: code slicing and retrieval mechanisms. The retrieval mechanism plays a vital role in performance enhancement, while setting a maximum loop iteration of 5 proves most efficient for error correction.
\end{tcolorbox}

\subsection{Practicability of \tool (RQ5)}
\subsubsection{Strategy}
We evaluate \tool's real-world performance on \datasetc (non-synthetic), benchmarking against SOTA LLMs using pass rate, BLEU-4, and error severity (ES) to assess accuracy and code quality.


\subsubsection{Result}

Table~\ref{tab:3-1} highlights \tool's significant performance boosts for LLMs on real-world Solidity data.



\begin{itemize}
    \item \textbf{Performance of various models:} Our approach integrates seamlessly with various LLMs, enhancing their real-world error correction for Solidity compilation issues. For instance, GPT-4o achieves a 96.97\% pass rate, outperforming its baseline (72.73\%) by 24.24\%. LLaMA3-8B improves from 45.46\% to 81.82\% (36.36\% of improvement) in pass rate. Deepseek-7B gains 30.30\% of improvement to reach 87.88\%, closing the gap between closed-source and open-source models. These results highlight our method's effectiveness in boosting small/medium open-source LLMs' performance, making them competitive for practical Solidity error correction.
    \item \textbf{Improvement of type-specific error:} 
    Our approach excels at fixing Solidity version migration errors, especially declaration/syntax errors (e.g., GPT-4o achieves 100\% correction). It also significantly improves complex error types. GPT-3.5-turbo shows a 39.39\% pass rate increase in type/parser errors, demonstrating \tool's capability to handle challenging real-world scenarios.
    This highlights \tool's adaptability across error categories and its value in enhancing LLMs' practical error correction accuracy.
\end{itemize}

\begin{tcolorbox}
\textbf{Answer to RQ5:} 
\tool is highly effective for resolving Solidity version-related compilation errors in real-world scenarios, narrowing the performance gap between proprietary and open-source models and significantly enhancing their overall error correction capabilities.
\end{tcolorbox}

\begin{table}[ht]
\centering
\caption{Performance in Real-World Dataset.}\label{tab:3-1}
\label{tab:1}
\small
\resizebox{0.49\textwidth}{!}{
\begin{tabular}{@{}c|c|c|c|c|c@{}}
\toprule
\textbf{Approach}               & \textbf{LLM}          & \textbf{Error Type} & \textbf{Pass Rate}  & \textbf{BLEU-4} & \textbf{ES}   \\ \midrule
\multirow{20}{*}{End-to-end LLMs} 
                            & \multirow{5}{*}{GPT-4o} & Parser        & 66.67\% & 97.01\%  & 95.75\% \\
                            &                        & Declaration  & 60.00\% & 94.77\%  & 99.58\% \\
                            &                        & Syntax         & 77.78\% & 93.08\%  & 98.24\% \\
                            &                        & Type     & 80.00\% & 92.15\%  & 92.96\% \\
                            \cmidrule(l){3-6}
                            &                        &{\textit{Average}}    & 72.73\% & 93.81\%  & 96.74\% \\ \cmidrule(l){2-6}
                            & \multirow{5}{*}{GPT-3.5-turbo} & Parser     & 33.33\% & 90.38\%  & 93.11\% \\
                            &                        & Declaration    & 60.00\% & 92.83\%  & 94.39\% \\
                            &                        & Syntax        & 44.44\% & 91.35\%  & 93.47\% \\
                            &                        & Type          & 20.00\% & 90.03\%  & 92.85\% \\
                            \cmidrule(l){3-6}
                            &                        &{\textit{Average}}    & 36.37\% & 91.15\%  & 92.85\% \\ \cmidrule(l){2-6}
                            & \multirow{5}{*}{LLaMA3-8B} & Parser      & 33.33\% & 89.93\%  & 94.33\% \\
                            &                        & Declaration  & 80.00\% & 93.33\%  & 97.65\% \\
                            &                        & Syntax        & 44.44\% & 91.17\%  & 94.21\% \\
                            &                        & Type          & 40.00\% & 87.94\%  & 91.37\% \\
                            \cmidrule(l){3-6}
                            &                        &{\textit{Average}}    & 45.46\% & 90.59\%  & 94.39\% \\ \cmidrule(l){2-6}
                            & \multirow{5}{*}{Deepseek-7B} & Parser     & 55.56\% & 92.41\%  & 91.94\% \\
                            &                        & Declaration  & 80.00\% & 97.09\%  & 93.77\% \\
                            &                        & Syntax         & 66.67\% & 93.87\%  & 93.26\% \\
                            &                        & Type          & 40.00\% & 91.31\%  & 88.87\% \\
                            \cmidrule(l){3-6}
                            &                        &{\textit{Average}}    & 57.58\% & 93.67\%  & 91.96\% \\ \midrule
\multirow{20}{*}{SMCFIXER}         
                            & \multirow{5}{*}{GPT-4o} & Parser  & 100.00\% & 97.87\%  & 97.76\% \\
                            &                        & Declaration   & 100.00\% & 99.39\%  & 99.07\% \\
                            &                        & Syntax         & 100.00\% & 98.41\%  & 98.65\% \\
                            &                        & Type          & 90.00\% & 97.49\%  & 95.84\% \\
                            \cmidrule(l){3-6}
                            &                        &{\textit{Average}}    & \textbf{96.97\%($\uparrow$24.24\%)} & \textbf{98.79\%($\uparrow$4.98\%)}  & \textbf{97.83\%($\uparrow$1.09\%)} \\ \cmidrule(l){2-6}
                            & \multirow{5}{*}{GPT-3.5-turbo} & Parser  & 55.56\% & 98.54\%  & 95.84\% \\
                            &                        & Declaration   & 100.00\% & 98.09\%  & 95.34\% \\
                            &                        & Syntax       & 88.89\% & 96.24\%  & 94.82\% \\
                            &                        & Type          & 70.00\% & 95.78\%  & 91.69\% \\
                            \cmidrule(l){3-6}
                            &                        &{\textit{Average}}    & \textbf{75.76\%($\uparrow$39.39\%)} & \textbf{97.16\%($\uparrow$6.01\%)}  & \textbf{94.92\%($\uparrow$2.07\%)} \\ \cmidrule(l){2-6}
                            & \multirow{5}{*}{LLaMA3-8B} & Parser & 66.67\% & 94.43\%  & 95.64\% \\
                            &                        & Declaration   & 100.00\% & 98.86\%  & 97.51\% \\
                            &                        & Syntax       & 88.89\% & 96.91\%  & 96.61\% \\
                            &                        & Type          & 80.00\% & 92.23\%  & 93.51\% \\
                            \cmidrule(l){3-6}
                            &                        &{\textit{Average}}    & \textbf{81.82\%($\uparrow$36.36\%)} & \textbf{95.61\%($\uparrow$5.02\%)}  & \textbf{95.89\%($\uparrow$1.5\%)} \\ \cmidrule(l){2-6}
                            & \multirow{5}{*}{Deepseek-7B} & Parser & 77.78\% & 95.24\%  & 96.96\% \\
                            &                        & Declaration  & 100.00\% & 98.79\%  & 97.28\% \\
                            &                        & Syntax        & 100.00\% & 97.17\%  & 97.82\% \\
                            &                        & Type          & 80.00\% & 93.80\%  & 92.96\% \\
                            \cmidrule(l){3-6}
                            &                        &{\textit{Average}}    & \textbf{87.88\%($\uparrow$30.30\%)} & \textbf{96.25\%($\uparrow$2.58\%)}  & \textbf{96.45\%($\uparrow$4.49\%)} \\ \bottomrule
\end{tabular}
}
\end{table}

\section{Related Work}

LLMs~\cite{dubey2024llama,guo2024deepseek} have shown strong potential in software engineering tasks like code generation~\cite{gu2023llm}, debugging~\cite{lee2024unified}, and translation\cite{pan2023understanding}. While closed-source models (e.g., GPT-4o) generally outperform open-source alternatives (e.g., LLaMA3), our \tool framework bridges this gap through architectural optimizations rather than scaling data. By combining retrieval-augmented generation (RAG)~\cite{lewis2020retrieval} with domain-specific knowledge (e.g., Solidity documents), it reduces hallucinations~\cite{guu2020} and enables multi-step reasoning~\cite{borgeaud2022}. This approach empowers open-source LLMs to achieve closed-source-level performance in tasks like compilation error resolution~\cite{yuan2024transagent}, proving that efficient design and dynamic knowledge integration can compensate for resource constraints while improving interpretability~\cite{thorne2021}.

Solidity plays a crucial role in blockchain development, yet existing research focuses on security/optimization of smart contracts rather than version evolution~\cite{zheng2023turn, su2023defiwarder, destefanis2018smart, singh2020blockchain}. Two critical gaps exist: (1) backward compatibility with legacy systems~\cite{wang2021security, chen2021maintenance} and (2) version-specific tooling constraints~\cite{di2019survey}. Evolving syntax and semantics between Solidity versions often causes compilation errors, hindering developer productivity. Our framework systematically addresses these challenges by enabling seamless version adaptation while maintaining ecosystem compatibility.

\section{Threats to Validity}
\subsection{Internal Validity.}
The primary internal threat stems from the construction of the dataset, as there is limited research on the Solidity language in the APR field. We utilize LLMs to build the dataset, and based on prior works~\cite{lei2024autocoder, lee2023making, lee2022personachatgen, li2024data}, employing LLMs for constructing datasets for various downstream tasks has proven effective. The second threat is that the validity of our \tool framework is only demonstrated within the constructed dataset. Therefore, we have gathered real Solidity compilation issues from open-source platforms to better showcase the generality of our method.
\subsection{External Validity.}
The main threat to external validity arises from the fact that not all changes documented in the official Solidity documentation lead to compilation errors. Including all such changes in the knowledge base could add unnecessary retrieval information. Therefore, we have selectively filtered the changes documented in the Solidity documentation based on different types of errors to enhance the accuracy of retrieval. Additionally, the evolution between Solidity versions may also involve minor versions. In the future, we will focus on the evolution of minor versions to accommodate a more comprehensive version evolution.

\section{Conclusion}

We propose the first LLM-powered framework to fix compilation errors caused by Solidity version updates. Our method extracts breaking changes from Solidity’s official documents to build a knowledge base. The retriever retrieves relevant error-fixing knowledge, which guides the LLM to generate precise patches. Experiments show our tool efficiently handles major Solidity version upgrades. Future work will extend support to minor versions for broader applicability.

\section{Acknowledge}
This research was supported by the Fundamental Research Funds for the Central Universities 226-2025-00004, and the National Natural Science Foundation of China under Grant No.62302437.

\bibliographystyle{IEEEtran}
\bibliography{conference_101719}

\end{document}